*Ab initio* study of the density dependence of the Grüneisen parameter at  pressures up to 360 GPa

Umesh C. Roy and Subir K. Sarkar[*]

School of Physical Sciences

Jawaharlal Nehru University, New Delhi-110067, India

 (*)    Email: ssarkar@mail.jnu.ac.in

## Abstract

*Ab initio* calculations based on the Density Functional Theory are used to show that the Debye  frequency is a linear function of density to a high accuracy for several elemental solids at pressures (at least) up to  360 GPa. This implies that the ratio of density over the (Debye-frequency-based) vibrational Grüneisen parameter is a linear function of density in this region. Numerical data from first principles calculations for several systems at temperatures up to 2000K suggest that this is also true for the thermal Grüneisen parameter in the same range of pressure.  Our analytical form of the vibrational Grüneisen parameter is applied  to an implementation of the Lindemann's melting criterion  to obtain a simple extrapolation formula for the melting temperatures  of materials at higher densities. This  prediction is tested against available experimental and numerical data for several elemental solids.

# 1  Introduction

The concept of the Grüneisen parameter [1] was introduced as a measure of anharmonic effects in the vibrational properties of a crystal. There are two somewhat different versions [2,3] of this parameter: the vibrational Grüneisen parameter ($\Upsilon_{vib}$) and the thermal Grüneisen parameter ($\Upsilon_{th}$). These concepts were introduced long time ago [4] and are well defined at all temperatures and pressures. In recent times these ideas have been applied in the description of matter under extreme thermodynamic conditions. To illustrate this we note that conditions of simultaneous high temperature and high pressure ( as is present, for example, in the planetary interior [5-16]) cannot presently be created on a sustained basis in the laboratory. One important way of circumventing this problem has been to perform measurements under more modest conditions and then to extrapolate the results to more extreme situations. For example, melting temperatures at very high densities are often predicted [17-21], using the Lindemann melting criterion, by using such a procedure. However, this requires having enough confidence in the validity of the formula to be used over the entire range of densities – from the relatively low value where the melting temperature is known to the much higher value for which the melting temperature is sought to be predicted  This, in turn, implies that the ingredients of the extrapolation formula, such as the Grüneisen parameter, should also be known accurately as a function of density.  If, in addition, the functional form of this dependence [22-25] is simple enough it aids the construction of a simple analytic extrapolation formula for the melting temperature.

Unfortunately, the requirement of having accurate data cannot always be fulfilled on the basis of laboratory experiments. But nowadays this issue can be addressed to a large extent by numerical calculations based on the modern density functional theories of electronic structure calculation. In this paper we present the results of such a study of the Grüneisen parameter -- both at zero temperature and at finite temperature. We show that the density dependence of (a) the vibrational Grüneisen parameter at zero temperature, calculated by using the Debye frequency, and (b) the thermal Grüneisen parameter at a fixed temperature, can be described, to a high level of accuracy, by very simple functional forms in the range of pressures that we study (up to 360

GPa) . One consequence of this simplicity is that it is possible, after making suitable approximations standard in the literature, to obtain a simple analytic expression for the dependence of the melting temperature on the density. The values of the vibrational Grüneisen parameter and the melting temperature at a reference density appear as parameters in this expression. If the data on the vibrational Grüneisen parameter is not available but the melting temperature data is available for a lower range of densities such an equation can still be used as an extrapolation formula for the melting temperature to higher densities (for example, under conditions that obtain in the interior of the Earth) – after treating the reference vibrational Grüneisen parameter as a fitting parameter. This can be of direct relevance in understanding the issue of the composition of the inner core of the Earth [12,26-28] as well as the conditions prevalent at the boundary between the liquid outer core and the solid inner core [21,29-32].

In the present work we calculate the vibrational Grüneisen parameters at zero temperature by making use of the Debye frequency. To calculate the thermal Grüneisen parameters we use various elements of modern density functional theories (DFT) [33]. This includes the density functional perturbation theory [34] for calculating the phonon spectra. Finite temperature aspects of vibrations are handled via the quasi-harmonic approximation that assumes that the amplitude of vibration always stays small enough so that the quantum statistics of vibration can be described in terms of a non-interacting gas of phonons.

The systems for which we calculate the vibrational Grüneisen parameter ($\Upsilon_{vib}$) at zero temperature are gold, copper, magnetic-nickel, platinum, palladium and rhodium in the face-centered-cubic structure and osmium and non-magnetic iron in the hexagonal-close-packed structure. Calculations of the thermal Grüneisen parameter ($\Upsilon_{th}$) are substantially more resource-intensive. Hence our studies are limited to platinum, palladium, rhodium, and iridium at temperatures up to 2000K. In both the cases the basic criterion for the choice of a system for study was the stability of its crystal structure over the entire range of thermodynamic conditions studied [35-49]. The basic results of our studies are the following:

(i) We show that the ratio of the density over either $\Upsilon_{vib}$ (at zero temperature) or $\Upsilon_{th}$ (at a fixed non-zero temperature) is a linear function of density to a very good approximation over a large and useful range of densities (corresponding to pressures up to 360 GPa – approximately the pressure at the Earth's core).

(ii) Using the result in part (i) regarding $\Upsilon_{vib}$ and making some suitable approximations, we derive an analytic expression for the melting temperature as a function of density -- by following previous works that implements the Lindemann criterion [50] for melting. This prediction is tested for four materials using data available in the literature.

The plan of the paper is as follows: In Section 2 we provide the definition of $\Upsilon_{vib}$ that is used in the present paper to calculate it at zero temperature from the vibrational spectrum calculated via the use of the density functional perturbation theory. We also provide a summary of the methodology used in the *ab initio* calculation of $\Upsilon_{th}$. Section 3 describes the numerical and the analytical results in detail and section 4 contains a summary and discussion.

## 2  Methodology

## 2.1  The vibrational Grüneisen parameter

The vibrational Grüneisen parameter ($\gamma_{vib,j}$) for a particular mode labelled by 'j' is defined [1] as

$$\gamma_{vib,j} = \frac{\partial \ln \omega_j}{\partial \ln \rho} \qquad (1)$$

, where $\rho$ denotes the density and $\omega_j$ is the vibrational frequency of the j-th mode. This simple looking definition however hides several complexities of numerical implementation related to the fact that it involves the comparison of a physical quantity at different values of the thermodynamic parameters. This change of thermodynamic state also changes the dimensions (and sometimes the shape also) of the unit cell – leading to a continuous change in the wave vector for a particular mode (labelled by 'j' above). Added to these difficulties is the fact that the wave vector is a continuous variable that can be anywhere in the first Brillouin zone. The measurement of the mode Grüneisen parameter is thus quite a complicated matter from the point of view of both computer-based calculations [51] and laboratory experiments. Hence it is common to deal with a representative vibrational frequency, namely the Debye frequency, to define the vibrational Grüneisen parameter. This leads to our operational definition, quite common in the literature, of the vibrational Grüneisen parameter as

$$\gamma_{\text{vib}} = \frac{\partial \ln \omega_D}{\partial \ln \rho} \quad (2)$$

where $\omega_D$ is the Debye frequency and is defined as

$$\omega_D = \left(\frac{3 \int g(\omega) d\omega}{\alpha}\right)^{1/3} \quad (3)$$

, where $g(\omega)$ is the phonon density of states [The notation $\gamma_D$ is also used for this particular definition of the vibrational Grüneisen parameter]. Here $\alpha$ is the zero frequency limit of the ratio $g(\omega)/\omega^2$.

## 2.2 The thermal Grüneisen parameter

The thermal (also called 'thermodynamic') Grüneisen parameter is defined [2] as

$$\gamma_{\text{th}} = \frac{1}{c_V}\left(\frac{\partial P}{\partial T}\right)_V \quad (4)$$

, where P and T denote the pressure and the temperature, respectively, and $c_V$ is the ratio of the constant-volume specific heat to the volume of the sample. In the following we explain all the steps of how we compute this parameter for a crystal lattice with a face-centered-cubic (fcc) structure [since all our calculations are for such cases (Pt, Pd, Rh, and Ir)]. Equation (4) shows that we need to calculate the pressure P at a uniformly spaced grid of temperatures (around the target temperature T) for a given volume so that the value of $\left(\frac{\partial P}{\partial T}\right)_V$ can be calculated by numerical differentiation. Since the unit cell of an fcc lattice has a fixed shape and only one lattice parameter, it is determined uniquely by the value of the density. To calculate the pressure this unit cell is subjected to a distortion described by the Lagrangian strain parameter **η** . Up to first order in **η** the value of the Helmholtz free energy per unit mass F is given by [52]

$$\rho^* F(\boldsymbol{\eta}, T) \;=\; \rho^* F(\boldsymbol{0}, T) + \sigma_{ij}\eta_{ij} \qquad (5)$$

, where $\rho^*$ is the mass density of the undistorted lattice, $\sigma_{ij}$ are the components of the stress tensor and $\eta_{ij}$ are the elements of the Lagrangian strain parameter $\boldsymbol{\eta}$. Summation over repeated indices is implied in equation (5). For a cubic lattice system (like the fcc) $\sigma_{ij} = -P\delta_{ij}$. The $\boldsymbol{\eta}$ that we use to calculate P is diagonal and is given by $\eta_{ij} = \theta\delta_{ij}$. Then

$$\rho^* F(\boldsymbol{\eta}, T) \;=\; \rho^* F(\boldsymbol{0}, T) - 3P\theta \qquad (6)$$

, up to first order in $\theta$. Thus $P = -\frac{\rho^*}{3}(\partial F/\partial \theta)_{\theta=0}$. The derivative $(\partial F/\partial \theta)_{\theta=0}$ is calculated numerically after computing F at several (N) values of $\theta$ between $-\theta_{max}$ and $\theta_{max}$ (distributed symmetrically around $\theta = 0$). The values of the pair (N, $\theta_{max}$) are (9, 0.01) for Pt and Ir and (11, 0.0107) for Pd and Rh.

A given value of $\theta$ defines a particular crystal lattice. To calculate the Helmholtz free energy per unit cell (f) for this crystal, we use the decomposition [53-55]

$$f \;=\; e_0 + f_{el} + f_{ph} \qquad (7)$$

, where $e_0$ denotes the ground state electronic energy per unit cell for the static lattice defined by the particular value of $\theta$. The thermal electronic and the lattice vibrational contributions to f are represented by $f_{el}$ and $f_{ph}$, respectively. The expression for $f_{el}$ is $(U - TS_{el})$. Here U is the average electronic excitation energy per unit cell at the temperature T -- as calculated from the Fermi-Dirac distribution applied to the single electron energy levels computed by using the density functional theory. It can be written as

$$U \;=\; <E_{el}(T)> - <E_{el}(0)> \qquad (8)$$

, where

$$<E_{el}(T)> \;=\; \int \varepsilon n(\varepsilon)\chi(\varepsilon, T)d\varepsilon \qquad (9)$$

The Fermi occupation function $\chi(\varepsilon,T)$ is given by $[\exp\left(\frac{\varepsilon-\mu}{k_B T}\right) + 1]^{-1}$ and $n(\varepsilon)$ is the electronic density of states per unit cell at the energy $\varepsilon$. The chemical potential ($\mu$) is calculated numerically by equating the integral $\int n(\varepsilon)\chi(\varepsilon, T)d\varepsilon$ to the total number of valence electrons per unit cell. <$E_{el}(0)$> is calculated via graphical extrapolation (to T = 0) of the values of <$E_{el}(T)$> calculated at somewhat higher values of the temperature T. The electronic entropy per unit cell is given by the expression

$$S_{el} = -k_B \int n(\varepsilon)[\chi \ln\chi + (1-\chi)\ln(1-\chi)]d\varepsilon \qquad (10)$$

The phonon contribution $f_{ph}$ in equation (7) can be written, within the quasi-harmonic approximation, as

$$f_{ph} = k_B T \int g(\omega)\ln[2\sinh(\frac{h\omega}{4\pi k_B T})]d\omega \qquad (11)$$

, where $g(\omega)$ is the phonon density of states -- normalized via the requirement that $\int g(\omega)d\omega$ equals three times the number of atoms per unit cell.

The quantity $c_V$ in equation (4) is calculated as the ratio of the constant-volume specific heat per unit cell ($c_0$) divided by the volume of the unit cell. $c_0$ is a sum of contributions from electronic and phononic excitations and can be written as

$$c_0 = c_{0,el} + c_{0,ph} \qquad (12)$$

Here

$$c_{0,ph} = k_B \int g(\omega)\left(\frac{h\omega}{2\pi k_B T}\right)^2 \varphi(1+\varphi)d\omega \qquad (13)$$

, with $\varphi$ being equal to $[\exp\left(\frac{h\omega}{2\pi k_B T}\right) - 1]^{-1}$. The electronic contribution ($c_{0,el}$) is given by

$$c_{0,el} = \int [\frac{1}{k_B T}\frac{d\mu}{dT} + \frac{(\varepsilon-\mu)}{k_B T^2}]n(\varepsilon)\chi(1-\chi)d\varepsilon \qquad (14)$$

, where $\chi$ is again the Fermi occupation factor. The derivative $\frac{d\mu}{dT}$, appearing in the expression for $c_{0,el}$, is calculated by evaluating the expression

$$-\frac{1}{T}[\int(\varepsilon-\mu)n(\varepsilon)\chi(1-\chi)d\varepsilon]/[\int n(\varepsilon)\chi(1-\chi)d\varepsilon].$$

## 3 Results

## 3.1 Vibrational Grüneisen parameter at zero temperature

To calculate the vibrational spectrum first-principles electronic structure based calculations are done by using the Density Functional Theory (DFT) [33] and Density Functional Perturbation Theory (DFPT) [34] as implemented in the Quantum Espresso [56] package. Convergence with respect to the **k**-grid (**k** being the Bloch wave vector) and the energy cut-off is checked so that a target of about 1 *meV* for the total energy per unit cell may be achieved at each density. After these initial tests the highest values of the energy cut-off and the **k**-grid resolution, as found across all the densities (for a given material), are used for the actual production runs at all densities for a given material. For phonon frequency calculations the **q**-grid (**q** being the phonon wave vector) is adjusted so that a convergence of 1 $cm^{-1}$ may be realized. Table 1 provides a summary of all the important DFT-parameters.

Table 1: DFT parameters used for the calculation of the vibrational spectra.

| System | Energy Cutoff (Ry) | **k**-grid | **q**-grid | PP/PAW data set | Library |
|---|---|---|---|---|---|
| Au | 60 | 24×24×24 | 12×12×12 | Au.rel-pz-dn-rrkjus_psl.0.1.UPF | Quantum Espresso |
| Cu | 70 | 24×24×24 | 10×10×10 | Cu.pbe-n-van_ak.UPF | |
| Ni | 70 | 14×14×14 | 12×12×12 | Ni.pbe-nd-rrkjus.UPF | |
| Pd | 80 | 28x28x28 | 12x12x12 | Pd.pz-n-rrkjus_psl.0.2.2.UPF | |
| Pt | 70 | 28×28×28 | 10×10×10 | Pt.pbesol-n-kjpaw_psl.0.1.UPF | |
| Rh | 130 | 24×24×24 | 10×10×10 | Rh.pbesol-spn-kjpaw_psl.0.2.3.UPF | |
| Os | 70 | 24×24×20 | 8×8×6 | Os.pw-mt_fhi.UPF | |
| Hcp-Fe | 120 | 24×24×16 | 6×6×4 | Fe.pw91-sps-kjpaw_psl.0.2.1.UPF | SSSP* |

*Standard Solid-State Pseudopotentials

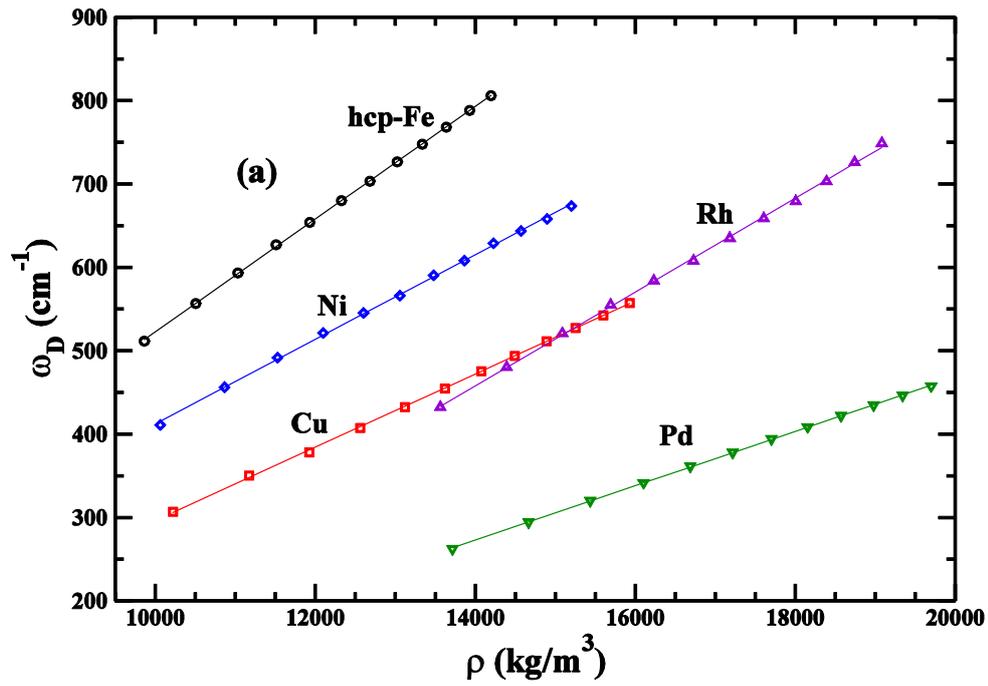

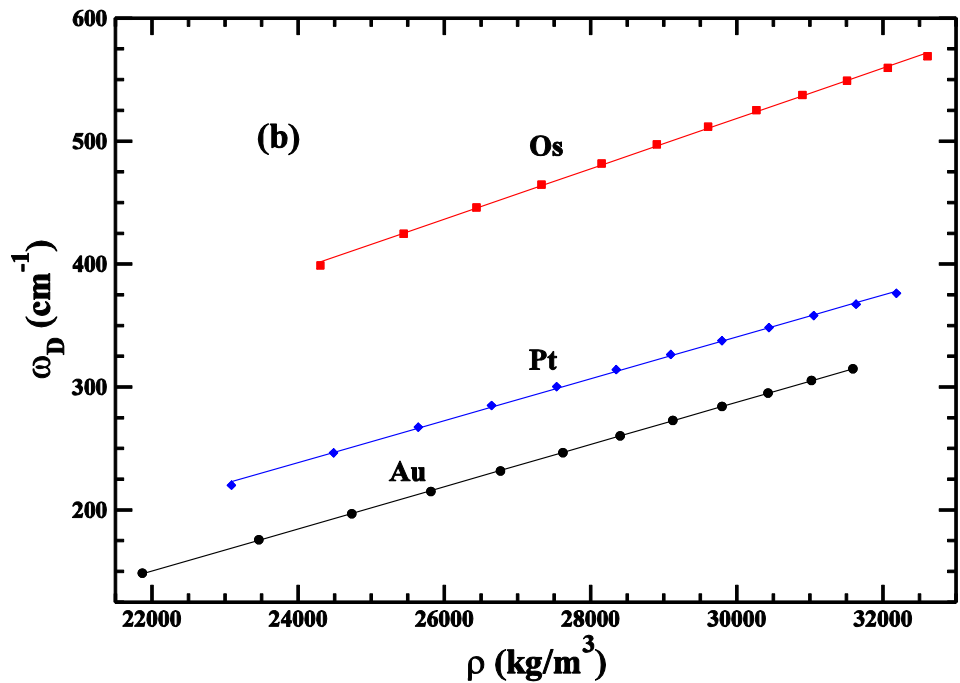

Figure 1: (Color online) Variation of the Debye frequency ($\omega_D$) with density ($\rho$) for (a) hcp-Fe, Ni, Cu, Rh and Pd; (b) Os, Pt and Au. The solid lines represent best fit straight lines.

Zero temperature calculations for all the elements have been done at pressures up to 360 GPa, at multiples of 30 GPa. This choice of the range of pressures takes care of the issue of the thermodynamic stability of the hexagonal-close-packed (hcp, non-magnetic) structure of iron [57-59].

The Debye frequency ($\omega_D$) is calculated in each case by fitting the functional form $(\alpha\omega^2 + \beta\omega^3)$ to the computed phonon density of states ($g(\omega)$) in the range of 0.04 $\omega_{max}$ to 0.16 $\omega_{max}$ -- where $\omega_{max}$ is the maximum value of the vibrational frequency for that particular spectrum. We find that this choice of the range of fitting provides stable value of the coefficient α (of equation (3)). Fig.1 shows a plot of the Debye frequency thus calculated against the density (ρ) for various elemental solids. We also show the plot of the best linear fit to the data for each material.

Even visually it is clear from fig.1 that the Debye frequency has a linear dependence on density to a high level of accuracy. Table 2 shows the parameters of this linear fit along with the correlation coefficients -- the linear fit being represented by the equation $\omega_D = a + b\rho$.

Table 2: Best fit parameter values and correlation coefficients from a linear regression analysis of the Debye frequency vs. density.

| System | a $(cm^{-1})$ | b $(\frac{cm^{-1}}{kg/m^3})$ | Nominal value of the correlation coefficient |
|---|---|---|---|
| Au | -227.419 | 0.017161 | 0.999973 |
| Cu | -142.832 | 0.043907 | 0.999878 |
| Ni | -94.992 | 0.050706 | 0.999663 |
| Pd | -183.024 | 0.032560 | 0.999936 |
| Pt | -170.522 | 0.017037 | 0.999597 |
| Rh | -330.603 | 0.056293 | 0.999655 |
| Os | -96.211 | 0.020482 | 0.999604 |
| Hcp-Fe | -153.655 | 0.067602 | 0.999947 |

The values of the correlation coefficients reflect the extent to which the linear relationship holds over the entire range of densities. If $\omega_D$ is indeed of the form (a +bρ), with a and b being material-specific constants, it immediately follows from equation (2) that the vibrational Grüneisen parameter at T = 0 will also have a correspondingly simple form. In fact

$$\frac{\rho}{\gamma_{vib}} = \frac{a}{b} + \rho \qquad (15)$$

Thus we have a rather simple relationship that **the ratio of density over the Debye vibrational Grüneisen parameter is a linear function of density with a slope of unity** (at least in this range of pressures). Since the value of $\gamma_{vib}$ is expected to depend primarily on density and rather weakly on temperature, it is reasonable to test the prediction made above against experimental data at room temperature.

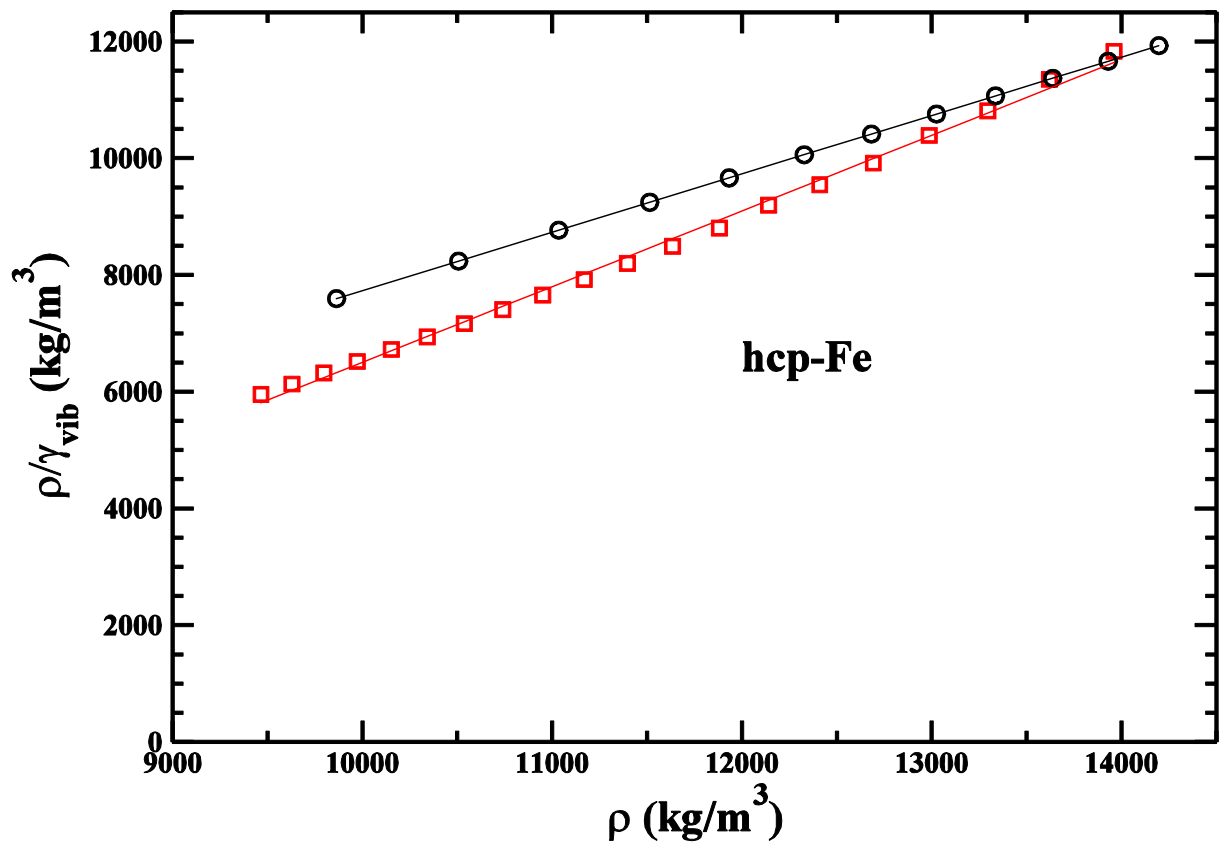

Fig.2: (Color online) Plot of the ratio of density over the vibrational Grüneisen parameter ($\rho/\gamma_{vib}$) against density ($\rho$) for hcp-iron: Circles, DFT data from the present study ; squares, experimental data [60,61]. The solid lines represent the best fit straight lines.

For hcp-iron such a comparison is shown in Fig.2. It can be seen that the prediction of linearity is supported quite well by the experimental data -- although the slope of about 1.3 is somewhat higher than the value of unity predicted by equation (15). This issue will be addressed in the Discussion section.

It is important to remember that the computed value of the ratio (a/b) will depend somewhat on the details of the implementation of the DFT. Hence it is useful to eliminate it and recast the equation (15) in the form

$$\gamma_{vib} = \frac{\rho \gamma_{vib,ref}}{\rho \gamma_{vib,ref} - \rho_{ref}(\gamma_{vib,ref} - 1)} \quad (16)$$

, where $\Upsilon_{vib,ref}$ is the vibrational Grüneisen parameter at some reference density $\rho_{ref}$. This equation gives us a tool for extrapolating the value of the vibrational Grüneisen parameter to higher densities from the knowledge of its value at a relatively low value of the density.

## 3.2 Thermal Grüneisen parameter

Table 3 shows the key parameters of the DFT-based calculations for the four elemental solids studied. The resolutions of the **k**-grid and the **q**-grid are such as to obtain a convergence of 1 *meV* for the Helmholtz free energy per unit cell.

Table 3: DFT parameters used for calculating the thermal Grüneisen parameter.

| System | Energy Cutoff (Ry) | **k**-grid | **q**-grid | PAW data set | Library |
|---|---|---|---|---|---|
| Pd | 80 | 24×24×24 | 4×4×4 | Pd.pbesol-n-kjpaw_psl.0.2.2.UPF | Quantum Espresso |
| Ir | 140 | 24×24×24 | 6×6×6 | Ir.pbesol-n-kjpaw_psl.0.2.3.UPF | |
| Pt | 80 | 28×28×28 | 6×6×6 | Pt.pbesol-n-kjpaw_psl.0.1.UPF | |
| Rh | 130 | 24×24×24 | 4×4×4 | Rh.pbesol-spn-kjpaw_psl.0.2.3.UPF | |

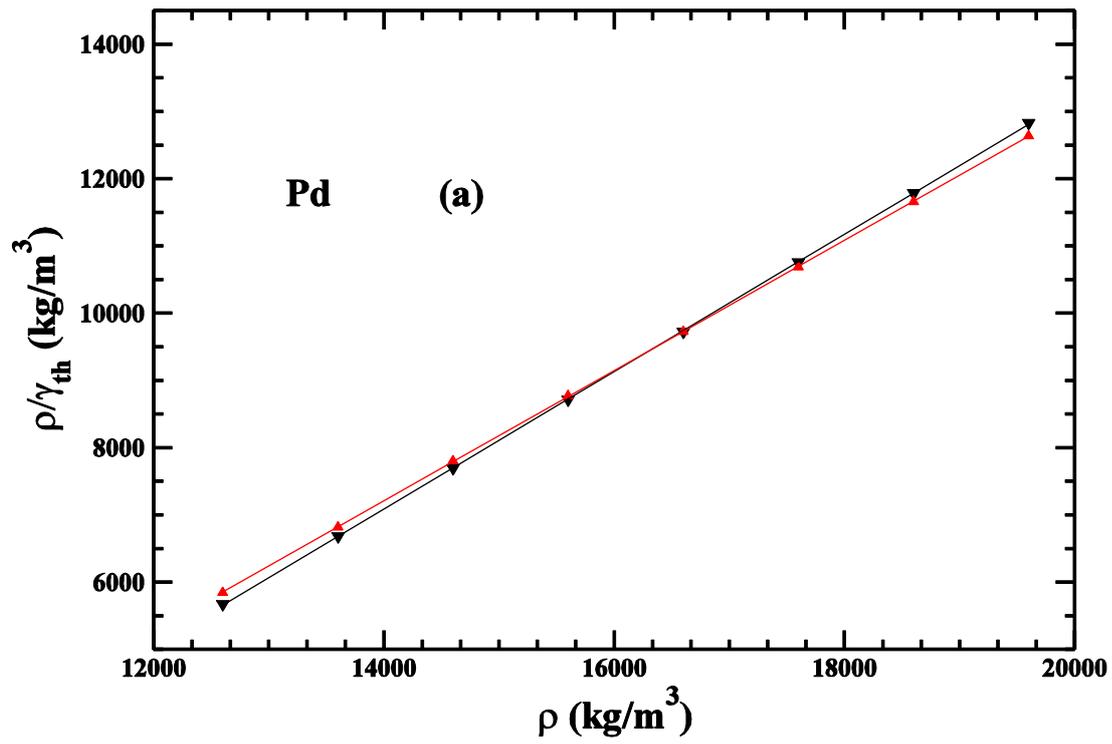

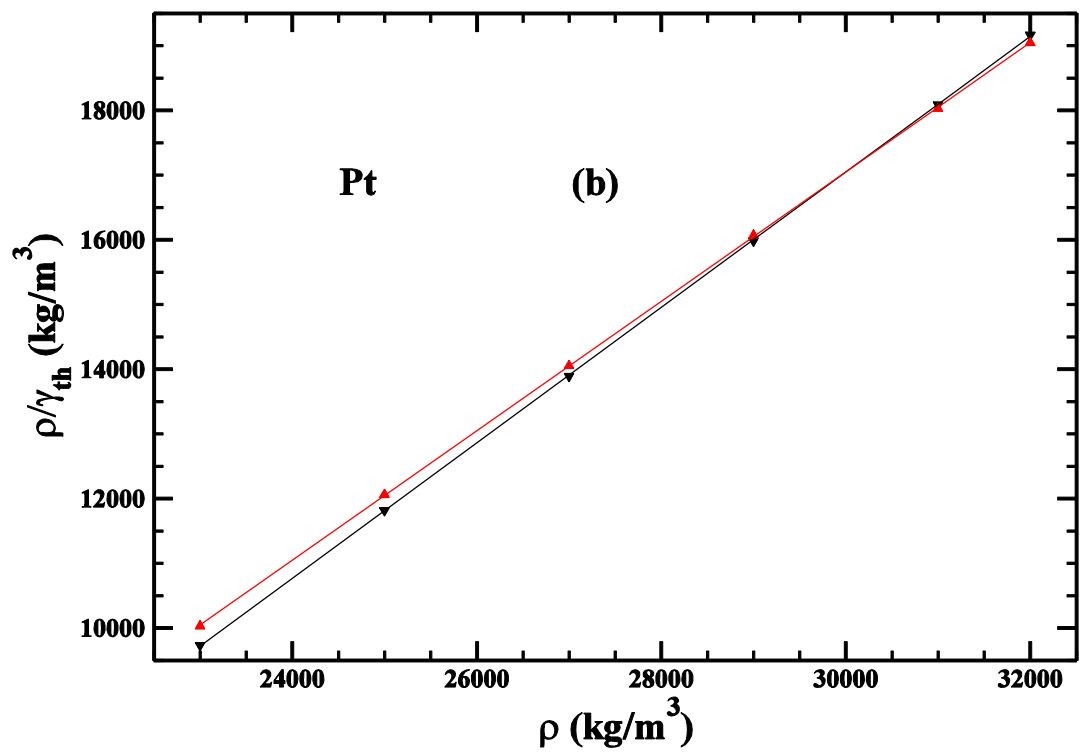

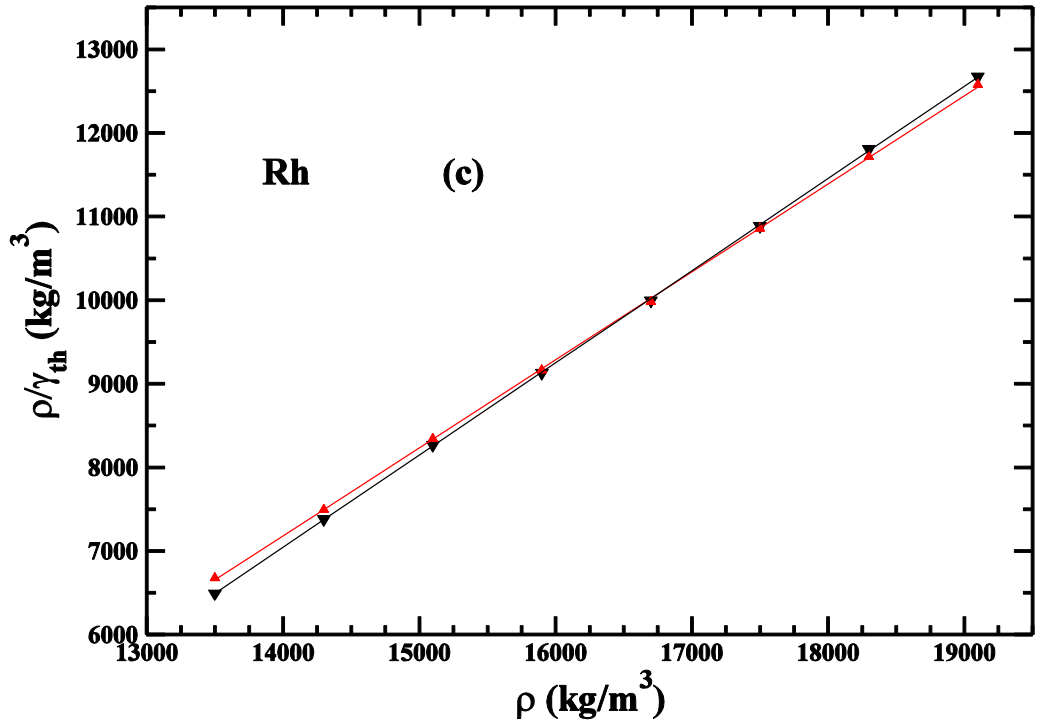

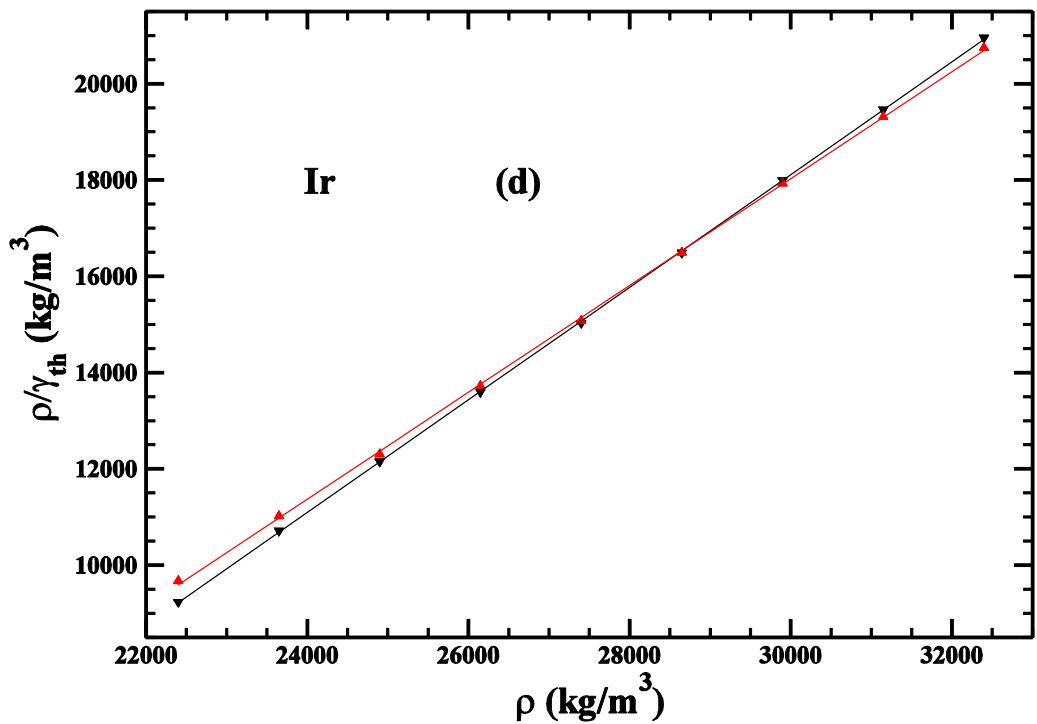

Fig.3: (Color online) Plots of the ratio of density (ρ) over the thermal Grüneisen parameter($\Upsilon_{th}$) against density for: (a) Pd, (b) Pt, (c) Rh, and (d) Ir [Filled inverted triangle, 300K ; Filled triangle, 2000K]. Straight lines represent the best fit to the data in each case.

Fig.3 shows the plots of the ratio of density ($\rho$) over the thermal Grüneisen parameter ($\Upsilon_{th}$) against density at two different temperatures for each material. The range of densities in each case is such that the pressure would vary approximately up to 360 *GPa* ( in the range of study variation of temperature causes only a minor change of pressure). The data for $\rho/\Upsilon_{th}$ is plotted at the temperatures of 300K and 2000K. It can be seen from the data that for all the systems studied temperature has a rather weak effect on $\Upsilon_{th}$ if the density is kept fixed. The effect of raising the temperature from 300K to 2000K on the value of $\Upsilon_{th}$ never exceeds five percent or so. Moreover, this slight variation has the same characteristics for all the four materials studied: with increasing temperature, $\Upsilon_{th}$ decreases slowly at the lowest densities but the trend is the opposite at the high density end. These facts provide precise quantitative support to arguments and statements made frequently in the literature on the weak temperature dependence of the thermal Grüneisen parameter.

The second prominent feature of Fig.3 is that, at a given temperature, $\rho/\Upsilon_{th}$ is a linear function of density to a very good degree – with a slope close to unity in all the four cases studied. Thus the observation of the linear dependence of $\rho/\Upsilon_{vib}$ on $\rho$ continues to hold good even when $\Upsilon_{vib}$ is replaced by $\Upsilon_{th}$ at a fixed temperature -- at least in the range of temperatures studied and for the range of densities studied. The upper limit imposed on the value of temperature in our calculations is dictated primarily by concerns about melting and a possible breakdown of the quasi-harmonic approximation.

Calculations on the density dependence of the thermal Grüneisen parameter, based on computational or experimental data, have been reported in the literature [55,62- 66]. However, the meaning of what is reported as $\Upsilon_{th}$ is not always the same. As a result comparison with previously reported results is not straightforward. In fig.4 we present two such (extracted and processed) sets of data (which we consider to be nearest in spirit to our calculations) in a format that permits a form of a test of the observation made earlier that the ratio $\rho/\Upsilon_{th}$ should be a linear function of $\rho$. The latter also implies that $1/\Upsilon_{th}$ should be a linear function of volume. This expectation is broadly fulfilled by the data sets presented in fig.4.

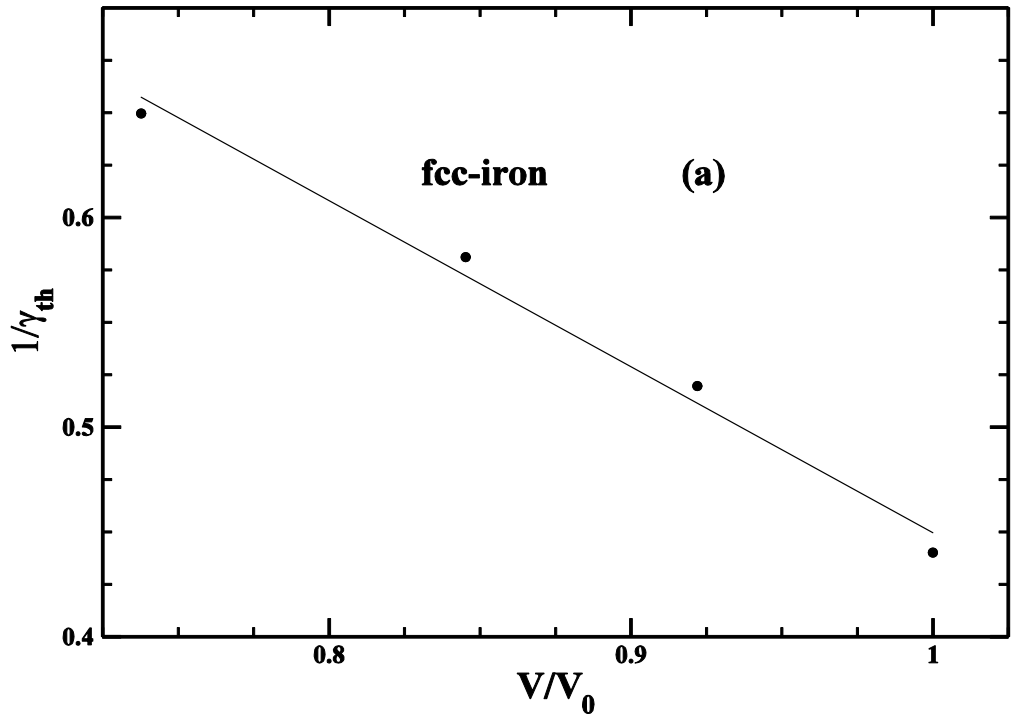

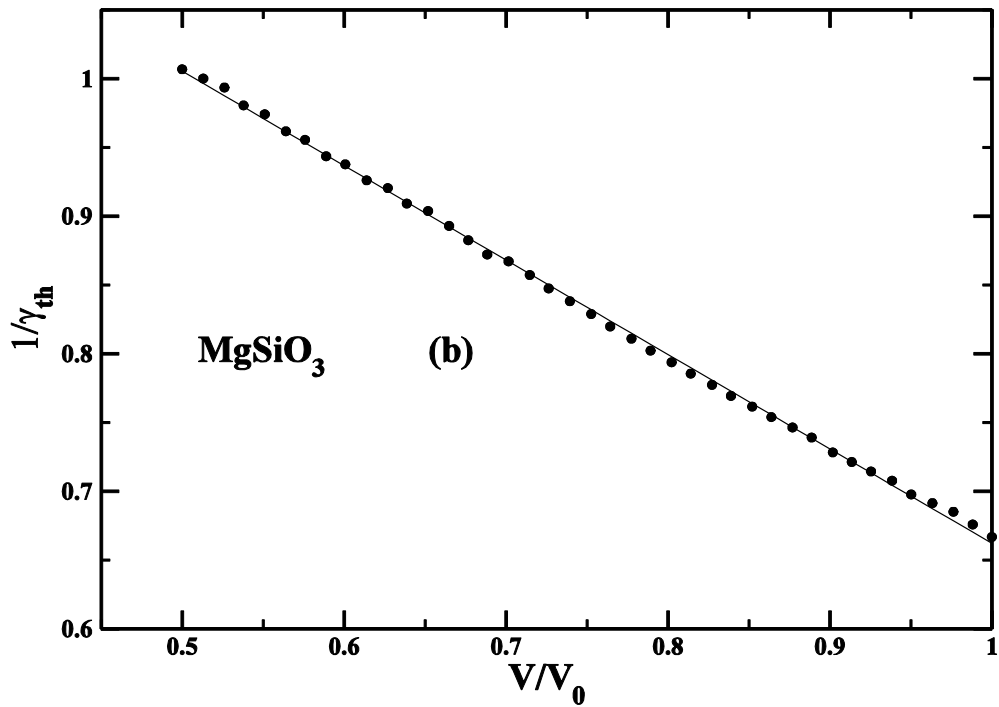

Fig.4: Inverse of the thermal Grüneisen parameter ($\gamma_{th}$) is plotted as a function of the scaled volume ($V/V_0$). $V_0$ is the volume at ambient pressure. (a) fcc-iron at 3000K [55] and (b) MgSiO$_3$ at 300K [66]. Also shown are the best fit straight lines.

## 3.3 Melting temperature at high densities

As an application of our observations regarding the density dependence of the vibrational Grüneisen parameter, we now visit the frequently discussed issue of how to extrapolate melting temperatures measured at lower densities to situations when the densities are much higher via the application of the Lindemann melting criterion. Several such formulas are available in the literature. The starting point for the derivation of all these formulas is the approximate equation (which can be derived in more than one way)

$$\frac{d\ln T_m}{d\ln \rho} = 2\left(\gamma_{vib} - \frac{1}{3}\right) \quad (17)$$

, where $T_m$ and $\gamma_{vib}$ are the melting temperature and the vibrational Grüneisen parameter, respectively, at the density $\rho$ [67-68]. It is on account of how the dependence of $\gamma_{vib}$ on the density $\rho$ is modelled that the various melting temperature extrapolation formulas differ. Typically these are of the form of power laws (in density) or linear combinations of multiple power law terms (including a constant) [22-25, 69]. We use the expression given by equation (16) – with the understanding that application should not be made, without further scrutiny, beyond the pressure domain in which it has been tested (i.e. not beyond 360 GPa). In conjunction with equation (17) this readily leads to the approximation formula:

$$\frac{T_m}{T_{m,ref}} = \left(\frac{\rho}{\rho_{ref}}\right)^{4/3} [\gamma_{vib,ref} - \frac{\rho_{ref}}{\rho}(\gamma_{vib,ref} - 1)]^2 \quad (18)$$

Here the qualifier `ref' means a reference state with, in practice, a relatively low density. Derivation of the equation (18) makes use of the assumption that the temperature dependence of $\gamma_{vib}$ is very weak and hence negligible in the entire range of thermodynamic parameters between the reference state and the situation where the melting temperature is sought to be calculated. This assumption is common to the alternative formulations also.

In order to test the validity of equation (18) we have extracted the melting temperature data (simulation and experiment) for four elemental solids (Cu, Pd, Pt and Au) from various sources in the literature. Conversion of the melting data from the {Pressure, Temperature} format to the

{Density,Temperature} format has been done by using the finite temperature Vinet equation of state [70]. The physical parameters required for this conversion are taken from [71] for Cu, Pt and Au and from [72] for Pd. We do a best fit to the scaled melting temperature ($T_m/T_{m,ref}$) vs. scaled density ($\rho/\rho_{ref}$) data via equation (18) -- with $\Upsilon_{vib,ref}$ being a fitting parameter. Fitting is done in each case by using only the experiment-based data. However, data from simulations are also shown for comparison. Fig.5 shows the plots of the extracted data and the best fit curves. Table 4 contains the values of the reference state densities and melting temperatures, as well as the best fit values of $\Upsilon_{vib,ref}$ and the value of $\Upsilon_{vib}$ predicted for the reference density by our *ab initio* calculations (equation (15) and table 2).

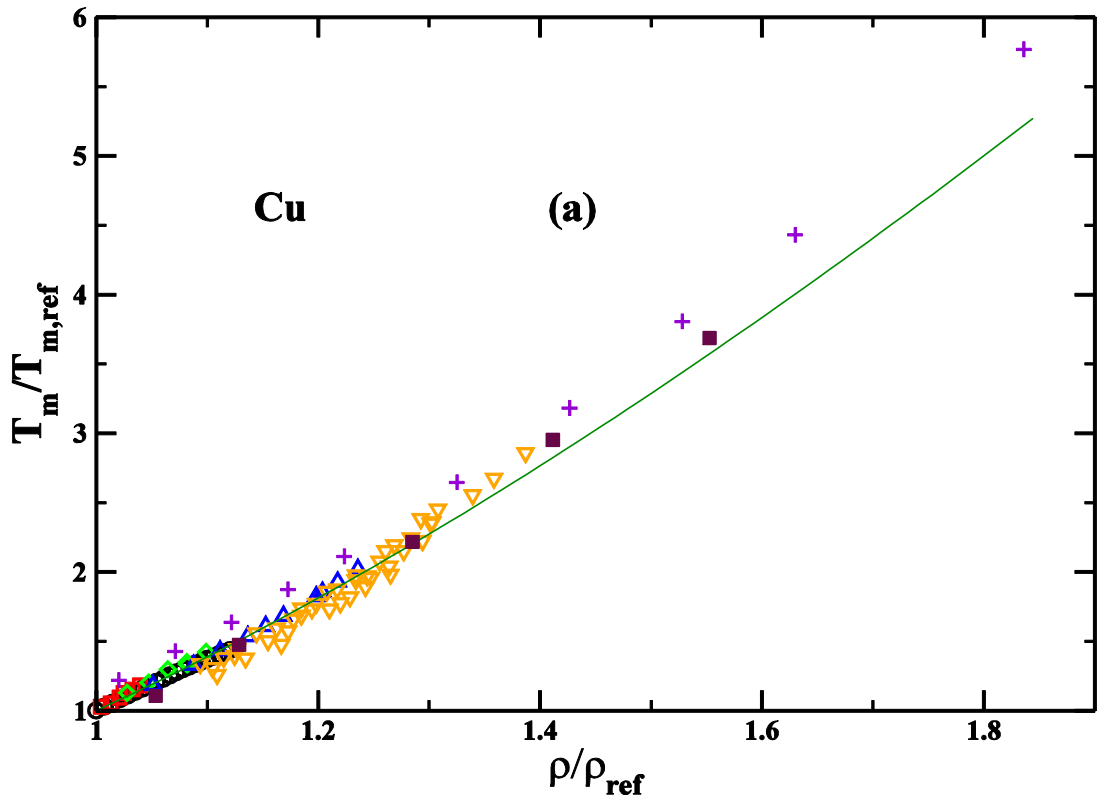

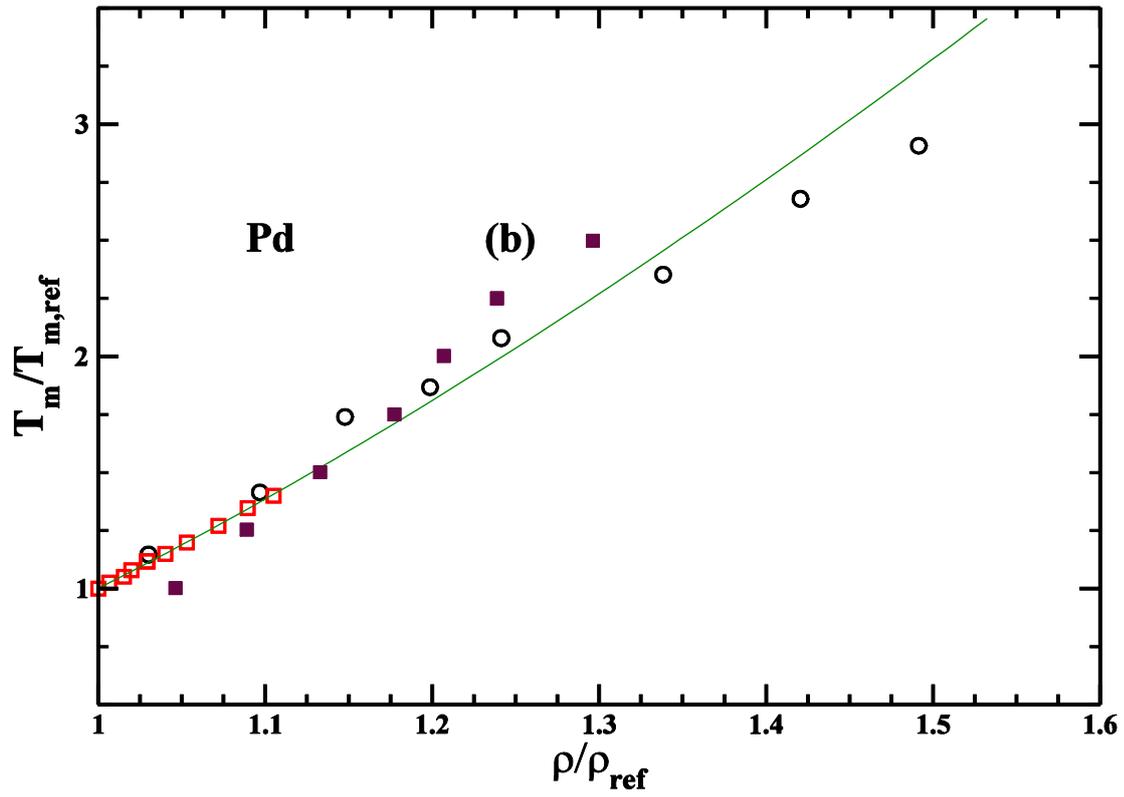

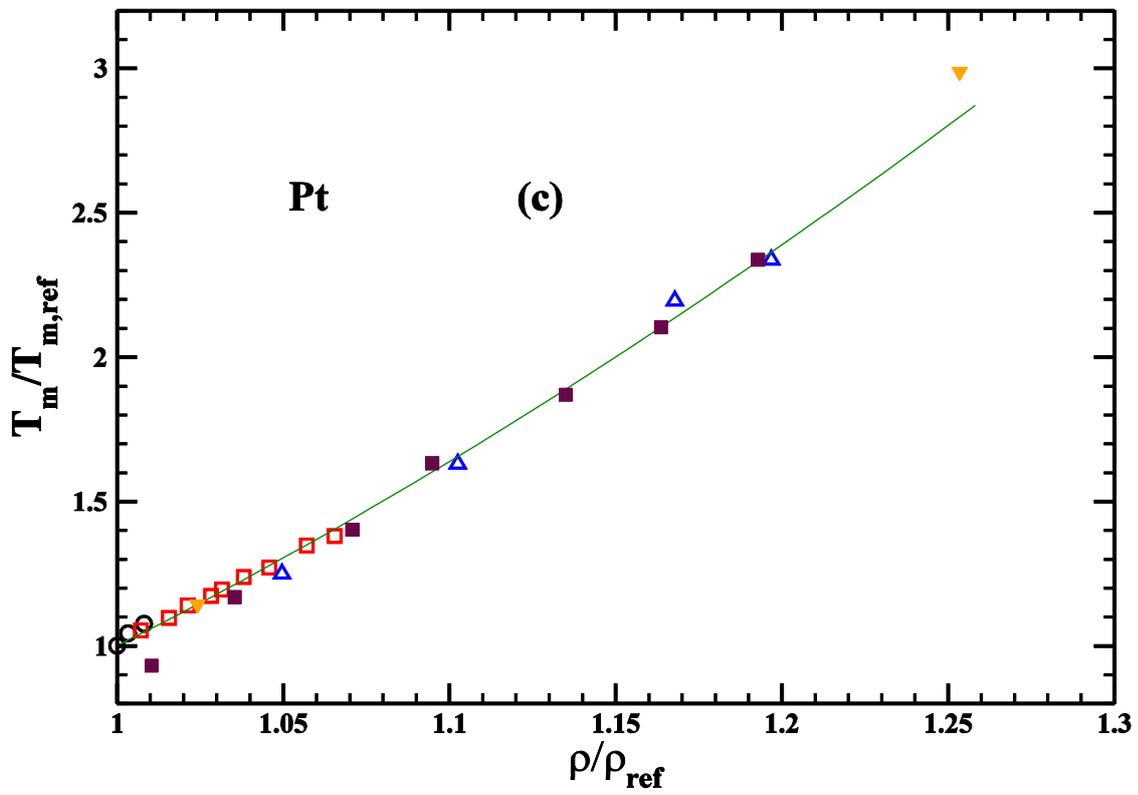

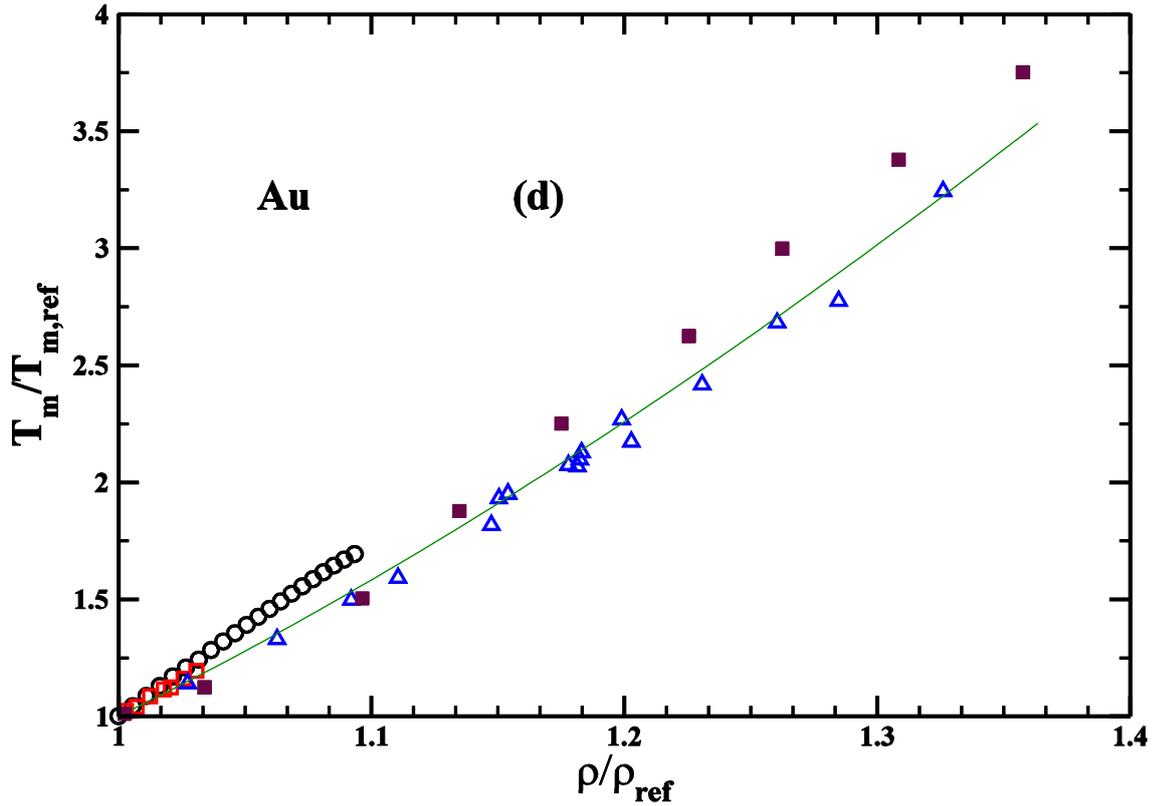

Fig.5: (Color online) Scaled melting temperature ($T_m/T_{m,ref}$) is plotted as a function of scaled density ($\rho/\rho_{,ref}$). (a) Cu: Cross [73], Circle[74], Open square[18], Diamond[75], Triangle[76], Inverted triangle[77], Filled square[78]; (b) Pd: Circle[73], Open square[76], Filled square[78]; (c) Pt: Circle[79], Open square[76], Triangle[80], Inverted filled triangle[81], Filled square[78]; (d) Au: Circle[74], Square[18], Triangle[82], Filled square[78]. Continuous lines represent the best fit curves given by equation (18) with $\Upsilon_{vib,ref}$ as a fitting parameter.

**Table 4:** Parameters of the reference state along with the best fit values and the predicted values of the reference Grüneisen parameter.

| Element | $\rho_{ref}$ $(\frac{kg}{m^3})$ | $T_{m,ref}$ (K) | $\gamma_{vib,ref}$ (Best fit value) | $\gamma_{vib}$ calculated at $\rho_{ref}$ |
|---|---|---|---|---|
| Cu | 8768 | 1356 | 2.15 | 1.60 |
| Pd | 12027 | 2002 | 2.15 | 1.88 |
| Pt | 21308 | 2140 | 3.21 | 1.89 |
| Au | 18988 | 1335 | 2.98 | 3.31 |

It may be noted that experimental data over a reasonably broad range of densities is available only for Au and Cu. For the other two cases this range is rather narrow. In fig.5 the nominal quality of fit of the extracted data to equation (18) varies from quite good to satisfactory . However, although the agreement between the best fit value of the reference vibrational Grüneisen parameter and that predicted by the DFT calculations at the reference density is quite good for Au and Pd it is of intermediate quality for Cu. For Pt the disagreement is substantial. While doing this comparison it is pertinent to note that the computed value of $\Upsilon_{vib,ref}$ will depend somewhat on the particular implementation of the density functional theory, the exact choice of the exchange-correlation energy functional etc.

## 4 Summary and discussion

The central result of this paper is the demonstration of the existence of a linear relationship between the density and the ratio of the density over the Grüneisen parameter for a set of elemental solids at pressures up to 360 GPa. For the thermal Grüneisen parameter this is true separately at each temperature. For the vibrational Grüneisen parameter this conclusion is based on the empirical observation of the linear dependence of the Debye frequency on density at zero temperature whereas, for the thermal Grüneisen parameter, this is based on its direct *ab initio* calculation. For the latter calculation the only restrictions that we need to impose are that the temperature should not be too close to the melting temperature and it should not be so high as to render the quasi-harmonic approximation inapplicable. The simple analytic expression of the vibrational Grüneisen parameter can be used, in conjunction with an implementation of the Lindemann melting criterion, to derive a correspondingly simple expression for the melting temperature as a function of density -- under the assumption of negligible dependence of the Grüneisen parameter on temperature.

The plot of the experimentally measured dependence of the ratio ($\rho/\Upsilon_{vib}$) on density for hcp-iron shows that it is linear to an excellent degree – as predicted by our DFT- based computations. However, the slope has a value of 1.3 – as opposed to the predicted value of unity. Resolution of this may lie in the way the Debye temperature is calculated in the experiment [60-61]. It is done on the basis of an analysis of the relationship between the experimentally measured value (via X-ray diffraction techniques) of the atomic mean square displacement and the Debye temperature – within the scope of the Debye approximation to the actual phonon density of states. The Debye approximation is accurate only at temperatures much lower than the Debye temperature. The experimental data that we examined was for room temperature. Thus the Debye approximation mentioned above is reasonably accurate only at the highest densities where the Debye temperature is also correspondingly high. However, as the density is reduced progressively the Debye temperature keeps going down and there is a steadily increasing violation of the requirement mentioned above. As a result we can expect a steadily increasing systematic error in the determination of the Debye temperature/frequency using the method employed in [60,61]. In figure 2 the agreement between the predicted and measured values of the ratio ($\rho/\Upsilon_{vib}$) is very good at the high density end -- as might be anticipated from the above argument. And, consistent with this same argument, the disagreement between the two keeps growing steadily as density is reduced. The question of whether the signature of this disagreement is also consistent with our reasoning requires a closer examination of the experimental data.

Agreement between the experimental data and our prediction for the melting temperature as a function of density presents a somewhat mixed picture. On the theoretical front one likely cause of this may be found in the derivation of equation (17) itself. A relatively transparent derivation of this equation [67] makes use of the Debye approximation to the actual vibrational density of states in the process of relating the mean square atomic displacement (and thus the Lindemann melting criterion) to the characteristic vibrational frequency (the Debye frequency) *at the melting temperature.* Such high temperatures will excite vibrational modes well beyond the low frequency region where the Debye approximation is reasonably accurate . This infirmity itself can substantially compromise the accuracy of equation (17).

**Acknowledgement**

We acknowledge the use of the High Performance Computing facility provided to the School of Physical Sciences, Jawaharlal Nehru University by the Department of Science and Technology (DST), Government of India under its DST-FIST-I and DST-FIST-II programs. UCR would like to acknowledge the fellowship provided by the University Grants Commission, India and additional financial support provided by the DST-PURSE program of the DST, Government of India.